\documentclass[twocolumn,prl,showpacs,aps]{revtex4}
\usepackage{graphicx}
\usepackage{dcolumn}

\newcommand{\pa}{\partial}
\newcommand{\td}{\tilde}

\newcommand{\beq}[1]{\begin{eqnarray}\label{#1}}
\newcommand{\eeq}{\end{eqnarray}}

\renewcommand{\footnote}{\footnote}

\begin{document}

\title{Holography and (1+1)-dimension non-relativistic Quantum Mechanics}

\author{Xiao-Jun Wang}
\email{wangxj@itp.ac.cn} \affiliation{Institute of Theoretical
Physics, Beijing, 100080, P.R.China}

\begin{abstract}
I generalize classical gravity/quantum gauge theory duality in
AdS/CFT correspondence to (1+1)-dimensional non-relativistic
quantum mechanical system. It is shown that (1+1)-dimensional
non-relativistic quantum mechanical system can be reproduced from
holographic projection of (2+1)-dimension classical gravity at
semiclassical limit. In this explanation every quantum path in
2-dimension corresponds to a classical path of 3-dimension gravity
under definite holographic projection. I consider free particle
and harmonic oscillator as two examples and find their dual
gravity description.
\end{abstract}


\maketitle

The holographic principle\cite{Hooft93,Susskind94} asserts that
the number of possible states of a region of space is same as
that of a system of binary degrees of freedom distributed on the
boundary of region. This principle has a realization in string
theory: Maldacena's conjecture on AdS/CFT
correspondence\cite{Mald98}. This conjecture states type IIB
string theory in AdS$_5\times S^5$ is dual to ${\cal N}=4$ super
Yang-Mills theory in four dimensions. In particular, it states
classical type IIB supergravity in AdS$_5\times S^5$ is dual to
large $N$ limit of quantum ${\cal N}=4$ super Yang-Mills in four
dimensions. It is well known that quantum mechanics seems to be
mysterious: In which the position and the momentum of a particle
should be interpreted statistically, or equivalently, in Feynman's
path integral language, a particle motioning between two different
space-time points will pass all possible paths or trajectories
simultaneously. These interpretations were abstracted into a few
quantum mechanics principles, such as uncertainty principle and
superposition principle, which is essentially different from
action principle of classical mechanics. However, in AdS/CFT
correspondence we see the duality between a classical theory and
a quantum theory. It indicates that, imposing the action
principle plus the holographic principle, a classical gravity
system can induce a quantum system which should be in general
governed by those quantum principles. Then a natural question is,
whether other quantum systems can be generated from a higher
dimension classical gravity system via imposing the holographic
principle? Or classical/quantum system duality in AdS/CFT
correspondence is only accident? The purpose of this letter is
just to provide some evidences to show that the classical/quantum
system duality may be a general result of the holographic
principle.

In order to achieve the above purpose, we should first understand
profound physical thought behind AdS/CFT correspondence. Here the
duality between classical supergravity and quantum conformal field
theory at large $N$ limit is exactly formalized by Witten's
conjecture\cite{Witten98}:
\begin{eqnarray}\label{1}
 W_{\rm gauge}[\phi_0]&=&-\ln<e^{i\int d^4x\phi_0(x){\cal
 O}(x)}>_{\rm CFT} \nonumber \\  &=&
 I_{\rm SUGRA}[\phi(z,x)=\phi_0(z_0,x)] \nonumber \\
 &=&\int_{-z_0}^{z_0}dz\int d^4x{\cal L}_{\rm SUGRA}[\phi(z,x)],
\end{eqnarray}
where $z_0\to\infty$ is cut-off at near boundary, $\phi(z,x)$ is
supergravity field and $z$ is holography dimension. That is, the
generator of connected Green's functions in the gauge theory, in
the large $N$ limit, is equal to the on-shell supergravity action.
When we compute correlation functions, in general it is valid to
take ansatz $\phi(z,x)=e^{ip\cdot x}Z(z)$. It means that for every
supersurface with $z=$constant, $\phi$ behaves as plane wave, but
$z_0$ as large radial cut-off to regularize divergence
corresponds to UV cut-off of quantum field theory. Then it is
clear that quantum effects come from propagation of $\phi$ in
holography dimension $z$. In other words, from Eq.(\ref{1}) we
can say that quantum corrections come from summing over all
values of $\phi$ along the classical path in $z$-direction.

Now let us consider a general, but simpler system: a particle
moving in (2+1)-dimension (denoting coordinates by $(t,x,z)$)
gravity field. Again, we assume that holographic screen is placed
at $z=z_0\to\infty$ and for sake of convenience coordinates on
this screen are still labeled by $(t,x)$. For convenient to
discussion we use a special holographic map that a point
$(t,x,z)$ is mapped to $(t,x,z_0)$. In other words, when a
particle moves from $(t',x',z')$ to $(t'',x'',z'')$, an observer
locating at the holographic screen will see that this particle
moves from $(t',x')$ to $(t'',x'')$\footnote{It is different from
AdS/CFT case that, here we use path/path correspondence to define
holographic projection, such that it is not necessary that the
endpoints of bulk path should be located at boundary.}.  Then a
classical path in (2+1) dimensions will be mapped to a classical
path in holographic screen. However, this map is not one-to-one.
Instead, many different classical bulk paths map to the same
classical path in holographic screen. Do all of these projected
paths be physical equivalent? In order to answer this question,
let us recall that the classical states describing the particle
moving in bulk can be labeled by $|x_{\rm cl}(t),z_{\rm cl}(t)>$,
and one in the screen is labeled by $|x_{\rm cl}(t)>$. However, it
is not true that $|x_{\rm cl}(t)>$ represents real physical state
in the screen, since obviously the information is lost in this
map. It conflicts to holographic principle which asserts the
information does not lose in this map. Therefore, we should label
the physical states in holographic screen by
 \beq{2} |\td{x}(t,z)>=|x(t),z(t)>,\eeq
instead of by $|x_{\rm cl}(t)>$. It implies that, at fixed $t$,
the infinite number of bulk states with $x=$const. (they fill a
$z$-axis) are one-to-one mapped to infinite number of boundary
states which fill $x$-axis. The parallel discussion can be
performed in momentum representation. That is,
 \beq{3} |\td{p}_x(t,p_z)>=|p_x(t),p_z(t)>.\eeq
The Eq.~(\ref{3}), however, induces an additional constrain that
holographic projection of all parallel paths in bulk are physical
equivalent since all of these states possess same $p_z(t)$.
According to the above naive discussion, we can suggest that
$(1+1)$-dimension single particle propagators $K(t',x';t'',x'')$
defined by Feynman path integral should associate classical
$(2+1)$-dimension gravity in terms of the following formula
 \beq{4} K(t',x';t'',x'')&\equiv& \int {\cal D}x(t)
 e^{\frac{i}{\hbar}\int_{t'}^{t''} dtL_{\rm QM}[x(t),\dot{x}(t)]}
 \nonumber \\
 &=&C\sum_{\rm {all\ possible\atop z(t'), z(t'')=z_0}}e^{iS_{\rm
 gr}/\hbar}, \eeq
where $L_{\rm QM}[x(t),\dot{x}(t)]$ is Lagrangian describing a
quantum mechanics system of single particle, $S_{\rm gr}=-m\int
ds$ is the action describing a particle moving in gravity fields,
and $C$ is normalization constant. In Eq.~(\ref{4}) we have
identified all parallel bulk paths by one whose one endpoint
locating at holographic screen (named $z(t'')=z_0$). This
identity means that every quantum path in holographic screen
corresponds to a classical path in bulk which is determined by
its initial value $z(t')$. This is essential idea of classical
gravity/quantum mechanics duality under holographic principle. It
should be noticed that the appearance of $\hbar$ in Eq.~(\ref{4})
is due to requirement of dimensionless but not one of
quantization.

The essential ideas of both Eq.~(\ref{1}) and Eq.~(\ref{4}) are
same. That is, at definite limit the quantum system can be induced
via summing over redundant information after projecting a higher
dimensional classical gravity to lower dimensional. The
differences are that Eq.~(\ref{1}) deal with a system with
infinite degrees of freedom, but Eq.~(\ref{4}) deal with finite
one. Both of them should be treated as corollary of holographic
principle. Naturally, we can conjecture that this idea is valid
for general $(1+1)$-dimension quantum systems, i.e., every
$(1+1)$-dimension quantum system has its dual gravity description.
In the following, we will consider two examples:
$(1+1)$-dimension non-relativistic free particle and harmonic
oscillator.

A particle moving in pure gravity field is described by geodesic
line equation,
 \beq{5} \frac{d^2x^\mu}{ds^2}+\Gamma^{\mu}_{\nu\lambda}
  \frac{dx^\nu}{ds}\frac{dx^\lambda}{ds}=0. \eeq
Conveniently we can take a diagonal metric ansatz,
 \beq{6} ds^2=f(z)(dt^2-dx^2)-g(z)dz^2. \eeq
It should be expected that the holographic screen is asymptotic
Minkowski such that $\lim_{z\to\infty}f(z)\to 1$. Then performing
first integral on equation of geodesic line, we have
 \beq{7} \dot{x}&=&v_x, \nonumber \\
         \dot{z}&=&\sqrt{\frac{f}{g}(v_z^2+(1-v_x^2)\ln{f})},
 \eeq
where dot denotes derivative to $t$, $v_x$ and $v_z$ are integral
constants. To integrate the second equation of (\ref{7}), another
integral constant is induced, which will be fixed by constrain
$z(t'')=z_0$. Therefore, summing over initial value $z(t')$ in
Eq.~(\ref{4}) can be transfer to sum over all possible $v_z$ or
$p_z=mv_z$,
 \beq{8} K(t',x';t'',x'')=\int\frac{dp_z}{2\pi\hbar} e^{iS_{\rm
 gr}/\hbar}, \eeq
where integrating over $p_z$ is usual integral instead of path
integral, and normalization constant $1/2\pi\hbar$ is determined
as like as usual quantum mechanics discussion. At non-relativistic
limit, the gravity action $S_{\rm gr}$ is approach to
 \beq{9} S&\simeq& -m\int dt+\frac{m}{2}\int dt
  (\dot{x}^2+\dot{z}^2) \nonumber \\
  &\simeq&\frac{1}{2m}\int dt [p_x^2+\frac{f}{g}
    (p_z^2+(m^2-p_x^2)\ln{f})]. \eeq
In the second line of the above equation we ignore the term
$m\int dt$ since it is independent of dynamics. If we take
$f(z)=g(z)=1$, from Eqs.~(\ref{8}) and (\ref{9}) we immediately
obtain
 \beq{10} K(t',x';t'',x'')=\sqrt{\frac{m}{2\pi i\hbar(t''-t')}}
 e^{\frac{im}{2\hbar}\frac{(x''-x')^2}{(t''-t')}}.
 \eeq
It is nothing but free particle propagator obtained by rigorous
calculation of Feynman's path integral. That is, if we
hologrphically project a classical free particle in flat
3-dimension space-time to 2-dimension, from viewpoint of a
2-dimension observer, the motion of this particle have to be
correctly described by quantum mechanics instead of classical
mechanics.

When we want to obtain propagator of $(1+1)$-dimension harmonic
oscillator from gravity, however, we will meet some unexpectedly
difficulties. The main reason is that the holographic map for this
gravity/quantum correspondence is no longer ``vertical'' but
should be ``oblique'', i.e., $(t,x,z)$ is no longer projected to
$(t,x,z_0)$ but in general should be projected to
$(t,\bar{x}(x,z),z_0)$. The natural conditions for this projection
are: 1) For every $z=$ constant, the map between $x$ and $\bar{x}$
should be one-to-one. 2) The holographic screen should be mapped
itself, i.e., $\pa\bar{x}/\pa x|_{z=z_0\to\infty}=1$,
$\pa\bar{x}/\pa z|_{z=z_0\to\infty}=0$. However, the above
conditions are not sufficient to define of holographic projection
uniquely. Therefore, it is obvious that the full dual gravity
description of $(1+1)$-dimension oscillator subtly depends on the
definition of holographic projection. Fortunately, it will be
shown that up to subleading order of $\hbar$ expansion, the dual
gravity description can be uniquely calculated even without
knowing any details of holographic projection.

The Feynman's propagator of harmonic oscillator yielded by
semiclassical method is
 \beq{11}K(t',\bar{x}';t'',\bar{x}'')=\sqrt{\frac{m\omega}{2\pi
 i\hbar\sin{\omega T}}}e^{iS_{\rm cl}/\hbar}, \eeq
with $S_{\rm cl}=\frac{1}{2}mA^2\omega^2\int_0^T dt\ \cos{2\omega
t}$, where for convenience we have taken $t'=0,\ t''=T$ and
$\bar{x}(0)=0$. The above propagator can be exactly reproduced by
means of Eq.~(\ref{8}) via
\begin{widetext}
 \beq{12}K(t',\bar{x}';t'',\bar{x}'')=\int\frac{dp_z}{2\pi\hbar}
   \exp{[\frac{im}{2\hbar}\int_0^Tdt\ (A^2\omega^2\cos{2\omega t}
   +\frac{p_z^2}{m^2}\cos{\omega t})]}. \eeq
\end{widetext}
Recalling we are now considering non-relativistic limit, up to
$O(p_z^4)$ we can require the second line of Eq.~(\ref{9}) to be
equal to exponential factor of r.h.d of Eq.~(\ref{12}), i.e.,
 \beq{13}&&v_x^2+\frac{f}{g}(v_z^2+(1-v_x^2)\ln{f}) \nonumber \\
    &=&A^2\omega^2\cos{2\omega t}+v_z^2\cos{\omega t}. \eeq
Consequently, we can solve $f(z)$ and $g(z)$ to obtain the dual
gravity description, at semiclassical approximation at least.
From the second equation of Eq.~(\ref{7}) we have
 \beq{14}t-t_0&=&\int\frac{dz}{\sqrt{1-v_x^2}}
  \sqrt{\frac{g}{f\ln{f}}}\left(1-\frac{v_z^2}{2(1-v_x^2)\ln{f}}
  \right) \nonumber \\ &&+O(v_z^4), \eeq
where $t_0$ is integral constant and we can conveniently set
$t_0=0$ due to the reason mentioned previous. Inserting
Eq.~(\ref{14}) to r.h.d of Eq.~(\ref{13}) we have
 \beq{15} &&A^2\omega^2\cos{2\omega t}+v_z^2\cos{\omega t}
  \nonumber \\ &=& A^2\omega^2\cos{[2\omega H(z)]}
  +v_z^2\cos{[\omega H(z)]} \nonumber \\
  &&-\frac{A^2\omega^2 v_z^2}{(1-v_x^2)^{3/2}}\cos{[2\omega H(z)]}
  \int\frac{dz}{\ln{f}}\sqrt{\frac{g}{f\ln{f}}} \nonumber \\
  &&+O(v_z^4), \eeq
where
 \beq{16}H(z)=\int\frac{dz}{\sqrt{1-v_x^2}}
   \sqrt{\frac{g}{f\ln{f}}}.\eeq
Then we obtain
 \beq{17} &&\frac{\cos{2H(\td{z})}}{\chi}=v_x^2
 +\frac{1-v_x^2}{h(\td{z})}, \nonumber \\
 &&\frac{f}{g}=\cos{H(\td{z})}-\frac{\cos{2H(\td{z})}}{\chi}
 \int dz\frac{h(\td{z})}{\ln{f(\td{z})}}, \eeq
where $\td{z}=\omega z$, $\chi=1/A^2\omega^2$ and
$h(\td{z})=H'(\td{z})$. The simplest case is $v_x=0$\footnote{Now
$v_x$ is no longer velocity of oscillator along $x$-direction
because of ``oblique'' holographic projection. It should be
treated as integral constant only, and may be determined by some
physical requirements.}, in which Eq.~(\ref{17}) simplifies to
 \beq{18} \chi&=&H'^2(\td{z})\cos{2H},  \nonumber \\
 \frac{dk}{dH}+k&=&\frac{2\chi\cos{H}\ \sin{H}}{\cos^2{2H}}-
 \frac{\chi\sin{H}}{\cos{2H}}, \eeq
with $k(\td{z})=(\ln{f})^{-1}$. From this equation we can
unambiguously obtain $f(z)$ and $g(z)$ such that metrics of the
dual gravity. In fact, for any non-relativistic quantum system
with Lagrangian
 \beq{19}L(t)=\frac{m}{2}\dot{x}^2-V(x), \eeq
the above method is valid. It also means that the dual gravity
configuration is independent of the definition of holographic
projection, up to subleading order of $\hbar$ expansion at least.

In fact, the relationship between quantum mechanics and classical
gravity was revealed even far before suggestion of holographic
principle, in black hole
thermodynamics\cite{Bekenstein72,Hawking75}. In a long time,
however, the most of attention focused on quantization of gravity
and this relationship was neglected. This situation has been
changed due to recent developments of string theory. Now more and
more evidences show that this relationship should exist indeed,
such that AdS/CFT correspondence, UV/IR relation\cite{PP98} and
radial/energy-scale relation\cite{RE} in gravity/guage duality,
and evaluation of $\beta$-function of gauge theory from classical
supergravity\cite{ABCPZ01}. In this letter, I discussed what role
the holographic principle plays in classical gravity/quantum
mechanics duality, and generalized Witten's formula to general
(1+1)-dimension non-relativistic quantum system. I suggested a
definition of holography between $(1+1)$-dimension
non-relativistic quantum system and $(2+1)$-dimension classical
gravity. In terms of this definition, I found the gravity dual of
$(1+1)$-dimension non-relativistic free particle, and harmonic
oscillator at semiclassical case. It is well-known that path
integral formula is complete definition of quantum mechanics, and
path integral formula includes all important ingredients of
quantum mechanics, such as uncertainty principle, quantum
superposition principle, and interference behavior of quantum
mechanics. Naturally, therefore, I can conjecture that, {\sl at
definite limit, action principle plus holographic principle in
(2+1)-dimension gravity system is equivalent to principles of
quantum mechanics in (1+1) dimensions}, or {\sl (1+1)-dimension
non-relativistic quantum mechanics can be interpreted as
holographic projection of classical gravity in (2+1) dimensions}.
This conjecture also indicates that: A observer (A) living in two
dimensions has to accept ``mysterious'' possibility description
on microscopic world because his classical measure method is not
sufficient to obtain all information of matter. A observer (B)
living in three dimensions, however, will find that he can obtain
same information as one obtained by observer (A) even using
classical measure method only. Therefore, from viewpoint of three
dimension observer, it is not necessary to accept possibility
interpretation to describe world observed by lower dimension
observer.

\end{document}